\documentclass{elsart5p}

\usepackage{amsmath,color,epsf,eufrak,graphicx}
\usepackage{wasysym}

\def\a{\alpha}
\def\b{\beta}
\def\g{\gamma}
\def\d{\delta}

\def\vt{\vartheta}

\begin{document}

\begin{frontmatter}
 \title{On Poincar\'e gauge
    theory of gravity,\\ its equations of motion, and Gravity Probe B}
\author{Friedrich W.\ Hehl\corauthref{cor1}} 
\address[cor1]{Institute
  for Theoretical Physics, University of Cologne, 50923 K\"oln, Germany\\and\\
  Dept.\ of Physics and Astronomy,
  University of Missouri, Columbia, MO 65211, USA}
\ead{hehl@thp.uni-koeln.de}
\ead[url]{http://www.thp.uni-koeln.de/gravitation/}

\author{Yuri N.\ Obukhov\corauthref{cor2}}
\address[cor2]{Theoretical Physics Laboratory, Nuclear Safety
  Institute, Russian Academy of Sciences, B.\ Tulskaya 52, 115191
  Moscow, Russia} 
\ead{yo@thp.uni-koeln.de}

\author{Dirk Puetzfeld\corauthref{cor3}}
\address[cor3]{ZARM, University of Bremen, Am Fallturm, 28359 Bremen, Germany}
\ead{dirk.puetzfeld@zarm.uni-bremen.de}
\ead[url]{http://puetzfeld.org}

\begin{abstract}
  We discuss the structure of the Poincar\'e gauge theory of gravity
  (PG) that can be considered as the standard theory of gravity with
  torsion. We reconfirm that torsion, in the context of PG, couples
  only to the {\it elementary particle spin} and under no
  circumstances to the orbital angular momentum of test particles. We
  conclude that, unfortunately, the investigations of Mao et al.\
  (2007) and March et al.\ (2011)---who claimed a coupling of torsion
  to {\it orbital} angular momentum, in particular in the context of
  the Gravity Probe B (GPB) experiment---do not yield any information
  on torsion. {\it File GPBtorsionPLA10.tex, 15 May 2013.}
\end{abstract}

\end{frontmatter}

\newpage

\section{Introduction}

Ever since E.Cartan in the 1920s enriched the geometric framework of
general relativity (GR) by introducing a {\it torsion} of spacetime,
the question arose whether one could find a measurement technique for
detecting the presence of a torsion field. Mao et al.\ \cite{Mao}
claimed that the rotating quartz balls in the gyroscopes of the
GPB experiment \cite{GPB}, falling freely on an orbit
around the Earth, should ``feel'' the torsion. We
  emphasize that the GPB team of Everitt et al.\ never made such a
  claim; they were aware that GPB would not be able to sense torsion
  \cite{Everitt}. However, similar to Mao et al., March et al.\
\cite{March:2011} argue with the precession of the Moon and the
Mercury and extend later their considerations to the Lageos satellite.

A consistent theory of gravity with torsion emerged during the early
1960s as gauge theory of the Poincar\'e group, see the review in
\cite{Reader}. This Poincar\'e gauge theory of gravity incorporates as
simplest viable cases the Einstein-Cartan(-Sciama-Kibble) theory (EC),
the teleparallel equivalent GR$_{||}$ of GR, and GR itself. So far, PG
and, in particular, the existence of torsion have {\it not} been
experimentally confirmed.  However, PG is to be considered as the
standard theory of gravity with torsion because of its very convincing
gauge structure.

Since the early 1970s up to today, different groups have shown more or
less independently that torsion couples only to the {\it elementary
  particle spin} and under no circumstances to the orbital angular
momentum of test particles. This is established knowledge and we
reconfirm this conclusion by discussing the energy-momentum law of PG,
which has same form for all versions of PG. Therefore, we conclude
that, unfortunately, the investigations of Mao et al.\ and March et
al.\ do not yield any information on torsion.

\section{Torsion defined, spin of matter introduced}

Einstein's theory of gravitation, GR, was finally formulated in
1916. Already since this time, mathematicians and physicists, namely
Hessenberg, Levi-Civita, Weyl, Schouten, and Eddington, amongst
others, started to develop the geometrical concept of a (linear) {\it
  connection} $\Gamma$. This is a tool for the parallel displacement
of vectors in a differential manifold, in particular in 4-dimensional
spacetime. The final formulation of the connection was given by
E.Cartan in 1923/24. He defined the connection 1-form
$\Gamma_\a{}^\b=\Gamma_{i\a}{}^\b dx^i$ as a new fundamental
geometrical object (with $\a,\b,..$ as frame and $i,j,...$ as
coordinate indices, both running {}from $0$ to $3$); for the explicit
references and for the formalism, including the conventions, compare
\cite{Reader}, pp.17--21.

If the connection is expressed purely in coordinate components, then
the antisymmetric part of it is a tensor, Cartan's {\it torsion}
tensor,
\begin{equation}\label{tor}
  T_{ij}{}^k= \Gamma_{ij}{}^k- \Gamma_{ji}{}^k\equiv
  2\Gamma_{[ij]}{}^k
=- T_{ji}{}^k,
\end{equation}
with its 24 independent components. This is the tensor alluded to in
the title of our paper. Mao et al.\ \cite{Mao} wanted to sense torsion
by using the results of the Gravity Probe B experiment of Everitt et
al.\ \cite{GPB}; later, March et al.\ \cite{March:2011} tried to do the
same thing by using data of the Moon, of the Mercury, and of the
Lageos satellite. We will come back to this issue later.

In GR, the Riemannian connection is represented by the Christoffel
symbols $\widetilde{\Gamma}_{ij}{}^{k}:=\frac{1}{2}
g^{kl}(\partial_{i}g_{jl}+\partial_{j}g_{l i}-\partial_{l} g_{ij})$,
where $g_{ij}$ are the components of the metric tensor and
$\partial_i:={\partial}/{\partial x^i}$. The Riemannian connection is
symmetric, it is torsion-free, that is,
$\widetilde{T}_{ij}{}^k=0$. Massive test particles in GR move along
the geodesics of the Riemannian connection: 
\begin{equation}\label{geodesics}
  \frac{d^2x^k}{d\tau^2}+\widetilde{\Gamma}_{ij}{}^{k}\frac{dx^i}{d\tau}
  \frac{dx^j}{d\tau}=0\,.
\end{equation}

When Cartan extended the geometrical framework of GR by introducing a
torsion of spacetime, he was conscious of the fact that he also had to
use a more fine-grained description of matter than in GR. Instead of a
classical fluid, acting via a symmetric energy-momentum density
$\mathfrak{t}$, he suggested a Cosserat type fluid with an asymmetric
energy-momentum density $\mathfrak{T}$ and an intrinsic or spin
angular momentum density $\mathfrak{S}$, see \cite{Reader}, pages 21
and 103.

This conception has been developed even before the spin of the
electron was discovered. We recognize that the introduction of the
geometrical concept of a torsion goes hand in hand with ascribing to
matter, besides an energy-momentum density, a further dynamical
characteristics, namely a spin angular momentum density. In a
general-relativistic theory of gravity, torsion and spin are
interdependent.

This interdependence was clear to Cartan. However, because of an
unfounded assumption, see Sec.\ \ref{universal}, he was not able to formulate a
consistent theory of gravity with torsion.

\section{Poincar\'e gauge theory as standard torsion theory}

In the early 1960s, a consistent framework for a valid physical theory
of torsion was initiated by Sciama \cite{Sciama} and Kibble
\cite{Kibble}. It was conceived as a gauge theory of the Poincar\'e
group \cite{Kibble}, the semi-direct product of the {\it translations}
(4 parameters) and the {\it Lorentz rotations} (6 parameters). In
Minkowski spacetime, the Poincar\'e group acts rigidly
(``globally''). By means of the gauge procedure \`a la
Weyl-Yang-Mills, the Poincar\'e group is ``localized'', acts merely
locally. This is made possible by introducing 4 gauge potentials for
the translations and 6 gauge potentials for the Lorentz rotations. The
emerging theory is called {\it Poincar\'e gauge theory} of gravitation
(PG), see \cite{Reader}, Part B for details.

The arena of the PG is a {\it Riemann-Cartan} (RC) {\it spacetime}. It
is determined by a metric $g_{\a\b}$ (and its reciprocal $g^{\g\d}$),
an orthonormal coframe $\vt^\a=e_i{}^\a dx^i$, and a Lorentz
connection $\Gamma^{\a\b}:=g^{\a\g}\Gamma_\g{}^\b=
-\Gamma^{\b\a}=\Gamma_i{}^{\a\b}dx^i$. Having such a connection, we
can define a covariant exterior derivative $D$. For a RC-space, we
find $Dg_{\a\b}=0$ (vanishing nonmetricity).

The coframe $\vt^\a$ can be understood as translational gauge
potential and the Lorentz connection $\Gamma^{\a\b}=-\Gamma^{\b\a}$ as
rotational gauge potential. The corresponding gravitational field
strengths are torsion and curvature, respectively, which we find by
differentiation of the corresponding potentials:
\begin{eqnarray}\label{tor'}
  T^\a&:=&D\vt^\a =d\vartheta^\alpha+\Gamma_\b{}^\a
\wedge\vt^\b\,,\\ \label{curv}
  R^{\a\b}&:=&d\Gamma^{\a\b}-\Gamma^{\a\g}\wedge\Gamma_\g{}^\b=-R^{\b\a}\,.
\end{eqnarray}

\begin{center}
\includegraphics[height=6cm]{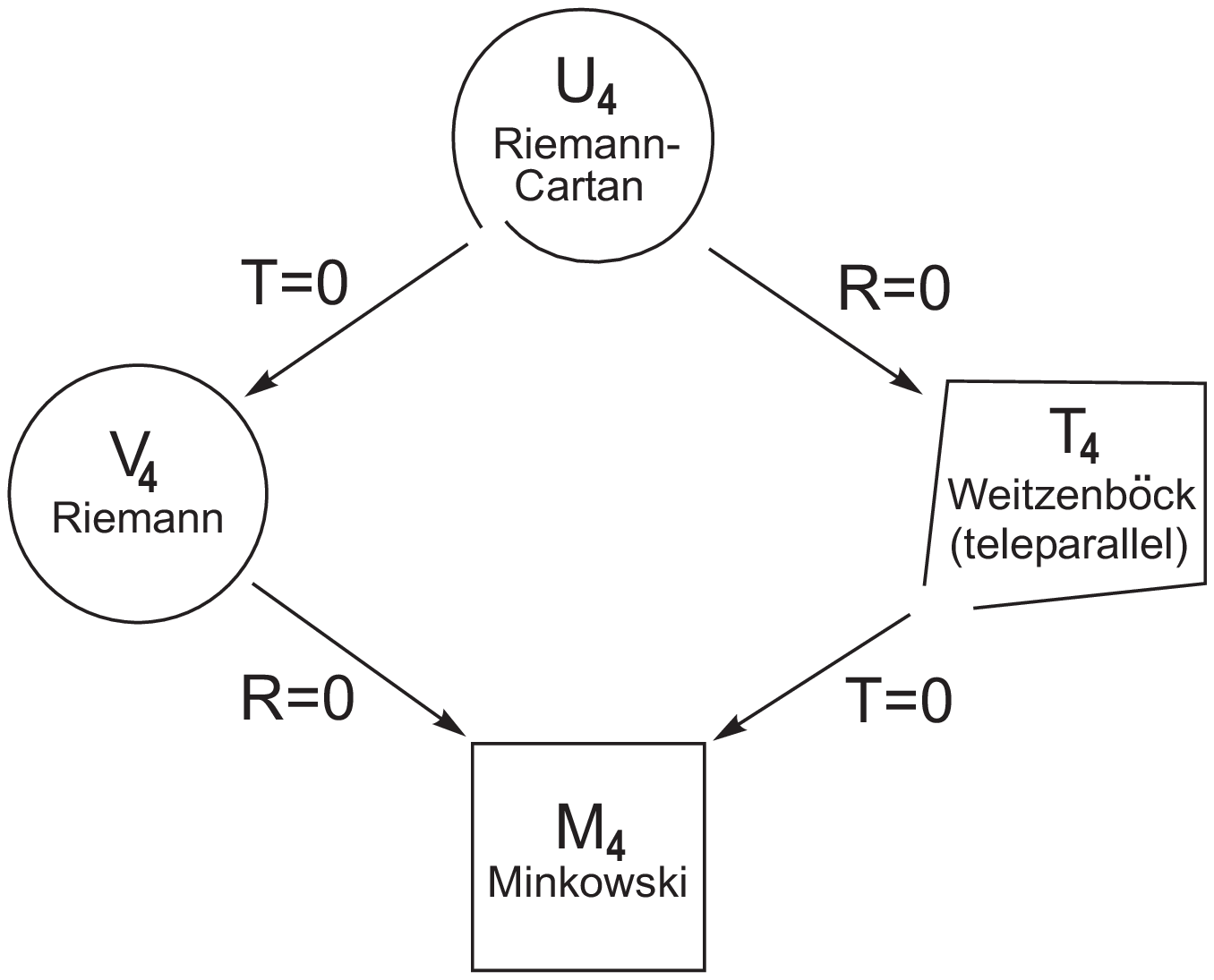}
\end{center}

\noindent{\it Fig.1.} A Riemann-Cartan space $U_4$ with torsion $T$ and
  curvature $R$ and its different limits (nonmetricity vanishes:
  $Q_{\a\b}:=-Dg_{\a\b}=0$), see \cite{Reader}, p.174.
\medskip

Note that in the term $\Gamma_\b{}^\a \wedge\vt^\b$ of (\ref{tor'})
the rotations and translation mix algebraically, due to the
semi-direct product structure. Hence it has to be taken with a grain
of salt that $T^\a$ is called the translation field strength. In
(\ref{curv}), the second term on the right-hand-side
$-\Gamma^{\a\g}\wedge\Gamma_\g{}^\b$ is due to the non-commutative
structure of the Lorentz rotations: they form a non-Abelian
sub-algebra of the Poincar\'e algebra.

The different limits of a RC space are represented in Fig.1. GR takes
place in a $V_4$, PG in a $U_4$, GR$_{||}$ in a $T_4$, and, when
gravity can be neglected, we are in an $M_4$.


The definition (\ref{tor'}) of the torsion, written with
respect to coordinates, degenerates to (\ref{tor}). Moreover, the
explicit form of the Lorentz connection, spelled out in coordinate
indices, is ${\Gamma}_{ij}{}^k=\widetilde{\Gamma}_{ij}{}^k
-K_{ij}{}^k$, with the contortion tensor
\begin{equation}\label{RCconn}
  K_{ij}{}^k=-{\scriptstyle\frac 12}(T_{ij}{}^k  
-T_{j\hspace{6pt}i}{}^{\hspace{-8pt}k}\;  +T^k{}_{ij})=
-K_{i\hspace{6pt}j}{}^{\hspace{-8pt}k}\;\,.
\end{equation}
So much about the geometry of the PG.

The physics of the PG is determined by a Lagrange 4-form
\begin{equation}\label{Lagr}
L=V(g_{\a\b},\vt^\a,T^\a,R^{\a\b})+L_{\text{mat}}(g_{\a\b},\vt^\a,\Psi,D\Psi)\,.
\end{equation} $V$ is the gravitational gauge part of the Lagrangian,
depending on the geometrical field variables, $L_{\text{mat}}$ is the
matter Lagrangian depending on some {\it minimally coupled} matter fields 
$\Psi(x)$, a Dirac field, for example. For special considerations
referring to nonminimal coupling, compare Sec.\ \ref{nonminimal}.

By varying with respect to the gauge potentials ($\d$ denotes a
variation), we can read off the sources in the field equations of the
PG as 
\begin{eqnarray}\label{sources}\mathfrak{T}_{\a}=\frac{\d
    L_{\text{mat}}}{\d\vt^\a}\quad\text{and}\quad
  \mathfrak{S}_{\a\b} =\frac{\d L_{\text{mat}}}{\d\Gamma^{\a\b}}=
  -\mathfrak{S}_{\b\a}\,,
\end{eqnarray}
respectively. They turn out to be the canonical 3-forms of {\it
  energy-momentum} $\mathfrak{T}_{\a}$ and of {\it spin angular
  momentum} $\mathfrak{S}_{\a\b}$ of matter.\footnote{They translate
  into the corresponding quantities of {\it tensor analysis\/} as
  follows: $\mathfrak{T}_\a={\cal T}_\a{}^\b\,\epsilon_\b$ and
  $\mathfrak{S}_{\a\b}={\cal S}_{\a\b}{}^\g\,\epsilon_\g$, with the
  3-form density $\epsilon_a:=e_\a\rfloor \epsilon$, the frame $e_\a$,
  and the volume 4-form density $\epsilon$. In the reverse order, we
  have $\vt^\b\wedge\mathfrak{T}_\a=\epsilon\,{\cal T}_\a{}^\b $ and
  $\vt^\g\wedge\mathfrak{S}_{\a\b}=\epsilon\,{\cal S}_{\a\b}{}^\g$.}

We postpone the discussion of the explicit form of the gravitational
Lagrangian $V$ since this is not necessary for the understanding of
the equations of motion of test particles in PG. We will only use it
later in order to see that PG embodies viable gravitational theories,
namely GR, Einstein-Cartan theory, and the teleparallel equivalent of
GR.

\section{How does one measure torsion of spacetime?}

We have now a general idea how a PG looks like. We recognize that PG
is a straightforward extension of GR, and we wonder, how a test
particle moves in a spacetime with torsion.

Clearly, we will take recourse to the established methods of GR. GR is
the only theory of nature in which the motion of a test particle is a
consequence of the field equation of that theory and of the
energy-momentum law following therefrom. Thus, $D\mathfrak{t}_\a=0$,
the energy-momentum law of matter in GR, yields the motion of the
momentum vector of a test particle, see, for instance, the textbook of
Papapetrou \cite{Papapetrou}, Chapter X (Equations of motion in
general relativity). The angular momentum law is trivial in GR. It
just entails the symmetry of the energy-momentum tensor of matter.

In PG, there emerges a second field equation of gravity and, induced
by it, an independent angular momentum law. Thus, in PG we have the
momentum and the angular momentum laws. This coupled set of equations
is used as a basis to derive equations of motion for test bodies. Why
should there be a reason to change horses that carried us so far in
the past? To postulate the equations of motion for a test particle
{\it independently} of the field equations controlling gravity is
potentially dangerous, because it will likely lead to inconsistencies,
and it defies established practice in GR.

In other words, the study of the equations of motion of test particles
in PG follows the same pattern as in GR. The only differences, see
\cite{Reader}, Chapter 14 (Equations of motion), are as follows: The
energy-momentum law picks up Lorentz type of forces,
\begin{equation}\label{energy}
  \partial_ j{\cal T}^{ i j}\cong\left(K_{ j k}{^ i}
    -\widetilde{\Gamma}_{ j k}{^ i}\right){\cal T}^{ j k}
  -{\cal S}^{ j kl}R_{ j k}{^ i}_l\, ,
\end{equation}
and the angular momentum law, for nonvanishing spin, becomes
nontrivial,
\begin{equation}\label{angular}
\partial_ k{\cal S}^{ i j k}
\cong{\cal T}^{[ i j]}+2\Gamma^{[ i}{}_{ kl}\,{\cal S}^{ j] kl}\,.
\end{equation}
These equations are here displayed in coordinate language for better
comparison with the older literature. They are ``weak identities'',
hence the $\cong$ sign, since we assumed the validity of the field
equation for matter: $\d L_{\text{mat}}/\d \Psi=0$.  We stress that
the two theorems (\ref{energy}) and (\ref{angular}) are generally
valid in PG independently of the explicit choice of the gravitational
Lagrangian $V$ in (\ref{Lagr}). That is, they apply to all torsion
theories that are formulated in a general-relativistic framework.

Let us now list chronologically a selection of decisive papers on the
measurement of torsion in order to provide an appropriate background
for the evaluation of the papers of Mao et al.\ \cite{Mao} and March
et al.\ \cite{March:2011}. In\medskip

\noindent$\bullet$ 1971: Ponomariev \cite{Pono} assumed that test
particles move along autoparallels (the straightest lines) of the
RC-spacetime:
\begin{equation}\label{auto}
  \frac{d^2x^k}{d\tau^2}+{\Gamma}_{ij}{}^{k}\frac{dx^i}{d\tau}
  \frac{dx^j}{d\tau}=0\,.
\end{equation}
There was no reason given. Of course, {}from a purely geometrical point
of view, these curves have a preferential role in a RC-space. This
does not imply, however, that they have to have a preferred role in
physics, too. We will see that this assumption reappears in the
literature later on. Then, in \medskip

\noindent$\bullet$ 1971, one of us \cite{Hehl:1971} pointed out that
``Torsion can be measured by means of a test particle with spin
possessing a canonical energy-momentum tensor with a nonvanishing
antisymmetric part.'' This was derived {}from the energy-momentum law
(\ref{energy}) of PG; note that the torsion enters (\ref{energy}) via
the contortion $K$, see (\ref{RCconn}). As far as we know, Ponomariev
did not object to this conclusion. In\medskip

\noindent$\bullet$ 1975, Adamowicz \& Trautman \cite{Adamowicz} took
the angular momentum law (\ref{angular}) and deduced the spin
precession induced by torsion: ``The torsion of space-time may be
measured by observing the precession of [the] spin of a particle.''
These two papers set the stage for the application of some more subtle
methods. In\medskip

\noindent$\bullet$ 1979, Rumpf \cite{Rumpf:1979}, in his Erice lecture
(given during the beginning of May 1979), computed by a
quantum-mechanical method the characteristic precession frequency of a
Dirac spin in a torsion field. This gave confidence that the spin
precession in a torsion field is a realistic effect, obeyed by one of
the fundamental fermionic fields of nature, see also
\cite{Rumpf:1981xh}. In \medskip

\noindent$\bullet$ 1979/80, Stoeger \& Yasskin \cite{Stoeger,Yasskin}
asked the question: ``Can a macroscopic gyroscope feel a
torsion?''. They used the general theory of spin motion of Mathisson
\& Papapetrou. Their verdict is unequivocal: ``Our results show that
the torsion couples to spin but not to rotation. Thus a rotating test
body with no net spin will ignore the torsion and move according to
the usual Papapetrou equations. Hence {\it the standard tests of
  gravity are insensitive to a torsion field} '' (emphasis by us). Is
there anything more to be said? This should have
been the (definite) end of the story. In \medskip

\noindent$\bullet$ 1981, Audretsch \cite{Audretsch:1981} considered
the Dirac electron in a spacetime with torsion. Since Rumpf
\cite{Rumpf:1979} had used a somewhat unconventional quantum-mechanical
procedure, it was reassuring that Audretsch, by employing a
WKB approximation of the Dirac equation in lowest order, found the
same precession frequency of the spin as Rumpf and the same effective
connection for the transport of the spin vector.

The basic understanding of the spin-torsion coupling was clarified at
this time (1981). For more details and further literature, we refer to
\cite{Reader}, Chapter 14. 
Still, let us have a quick look at some subsequent papers for
curiosity. In\medskip

\noindent$\bullet$ 1997, L\"ammerzahl \cite{Lammerzahl:1997wk}
revisited the Hughes-Drever experiment, which was originally used to
exclude a possible anisotropy of the mass. He determined the influence
of torsion on the energy levels of the atoms involved and found as
upper bound for the {\it axial} piece of the torsion $^{(3)}T^\a <
10^{-15}\, meter^{-1}$.

This result reminds us that the remaining irreducible pieces of the
torsion,\footnote{Explicitly, we have for the (co)vector and the axial
  (co)vector pieces ${\cal V}:=e_\b\rfloor T^\b$ and ${\cal
    A}:=\,^\star(\vt_\a\wedge T^\a)$, with $\,^{(2)}T^\a=-\frac 13
  {\cal V}\wedge\vt^\a$ and $\,^{(3)}T^\a=\frac 13 \,^\star({\cal
    A}\wedge \vt^\a)$, respectively, see \cite{Reader}, p.225.} namely
the tensor piece $^{(1)}T^\a$ and the vector piece $^{(2)}T^\a$, with
$T^\a=\,^{(1)}T^\a+\,^{(2)}T^\a+\,^{(3)}T^\a$, must be measured by
means of test particles with spins $s\ne\frac 12$. We extract {}from the
results of Seitz \cite{Seitz:1986} and of Spinosa
\cite{Spinosa:1987a,Spinosa:1987b} the following formula for the
torsion as seen by a test spin s, which is valid for $s=\frac 12,
1,\frac 32, 2$:
\begin{eqnarray}\label{torsionasseen} T^\alpha_{s\, >\, 0}&=&\bigl(1-\frac{1}{
    2s}\bigr)\,T^\alpha+\frac{3}{2s}\,^{(3)}T^\alpha\nonumber\\ &=&
  \bigl(1-\frac{1}{2s}\bigr)\bigl(\,^{(1)}T^\alpha+\,^{(2)}T^\alpha\bigr)+
  \bigl(1+\frac{1}{s}\bigr)\,^{(3)}T^\alpha\,. 
\end{eqnarray} 
Accordingly, a Proca field, which carries spin 1, couples, in contrast
to the Dirac field, to all three pieces of the torsion. Subsequently,
in 2008, Kostelecky, Russell, and Tasson \cite{Kostelecky:2007kx}, by
using new data, confirmed the upper bound for a possible torsion. In
\medskip

\noindent$\bullet$ 2000 Kleinert \cite{Kleinert:1998cz}, quite
surprisingly, ``proved'', neglecting almost all of the previous
literature on equations of motion, that a spinless particle follows an
autoparallel path thereby sensing torsion. This contradicts
established theories, see our discussion above of Ponomariev (1971)
and the consequences.

Kleinert takes an argument {}from particle physics. The spin 1 of a rho
vector-meson, if considered as a bound state on a quark level, may be
only caused by {\it orbital} angular momentum, that is, the spin 1 of
the $\rho(770)$ may be orbital angular momentum in
camouflage. However, this argument forgets the lesson of {\it
  effective field theory.}\footnote{``In physics, an {\it effective
    field theory} is, as any effective theory, an approximate theory,
  (usually a quantum field theory) that includes appropriate degrees
  of freedom to describe physical phenomena occurring at a chosen
  length scale, while ignoring substructure and degrees of freedom at
  shorter distances (or, equivalently, at higher energies)'', see {\it
    Wikipedia} of 30 March 2013.}

If a $\rho(770)$ moves at moderate speed in an exterior gravitational
field, there is every reason to believe that it does behave like a
Proca field of spin 1, in accordance with its classification in
elementary particle tables. In fact, we know experimentally in the
case of a neutron, moving in a gravitational field, that it behaves
like a Dirac particle of spin $\frac 12$, see the
Colella-Overhauser-Werner (COW) experiment \cite{COW} and its
interpretation \cite{RauchWerner,Alexandrov}. Whether the neutron spin
$\frac 12$ has, on the quark level, an orbital angular momentum
contribution of the three quarks, has no relevance for the COW-type
experiment. If we are in a quark-gluon plasma, however, then the quark
spin is of relevance and torsion couples to it, but under normal
conditions the neutron is of spin $\frac 12$.

Accordingly, Kleinert's new universality principle of orbital and spin
angular momentum mixes up different levels of observation. Above all,
Kleinert's conclusion on the autoparallels drawn therefrom, as we saw
above, defies all knowledge on equations of motion in
general-relativistic field theories. His nuclear physics arguments are
contrived and do not apply to the neutrons of the COW experiment nor
to the $\rho(770)$. In \medskip

\noindent$\bullet$ 2002 Shapiro \cite{Shapiro:2002}, in his extended
review of torsion, found again equations of motion for a spin in a
torsion field that are consistent with those of Rumpf, Audretsch, and
L\"ammerzahl mentioned above, see \cite{Shapiro:2002}, Eq.\ (4.60). However,
in\medskip

\noindent$\bullet$ 2007, Mao, Tegmark, Guth, Cabi\footnote{In 2006,
  when the Mao et al.\ paper was uploaded to arXiv.org, one of us
  immediately communicated his objections to Max Tegmark and his
  coauthors, basically the same objections as those to be discussed in
  this Letter; but it was of no avail. The analogous happened in the
  case of the March et al.\ papers. They did not find our arguments
  convincing either.} \cite{Mao}, in their investigation on the
possible effect of torsion on the rotating quartz balls of the
gyroscope of the Gravity Probe B experiment \cite{GPB}, proposed two
postulates: (i) The equation of motion for their ``spin'', see
\cite{Mao}, Eq.\ (19), and (ii) their spin has to move along an
autoparallel. Both postulates are ad hoc and the second one is even
inconsistent, as we saw above. We will come back to their paper in
Sec.\ \ref{answerMao}. Already before the final publication of the Mao
et al.\ paper, in\medskip

\noindent$\bullet$ 2007, Flanagan \& Rosenthal \cite{Flanagan:2007dc}
noted that the gravitational theory with torsion, taken by Mao et al.\
as a guinea pig, is inconsistent. This led Mao et al.\ to declare that
the torsion theory they used should ``...not be viewed as a viable
physical model, but as a pedagogical toy model giving concrete
illustrations of the various effects and constraints that we
discuss.'' Why should an inconsistent model be good enough for
pedagogical purposes if it is basically not good enough for a
scientific journal?

At the same time Flanagan \& Rosenthal \cite{Flanagan:2007dc} stated
in their conclusions that ``There may exist other torsion theories
which could be usefully constrained by GPB. It would be interesting to
find such theories.'' In Sec.\ \ref{answerMao} we will show that such
a finding appears to be only a very remote possibility. Shortly
afterwards, in\medskip

\noindent$\bullet$ 2007/08, Puetzfeld \& Obukhov
\cite{DirkYuri:2007,DirkYuri:2008} showed by a multipolar approximation
scheme that Mao et al.\ are ruled out for a very large class of
theories, only intrinsic spin couples to torsion, in particular it was
explicitly shown, see Sec.\ 10 of  \cite{DirkYuri:2008}, that the model
of Hayashi and Shirafuji \cite{Hayashi:Shirafuji:1979} does {\it not}
have the properties claimed by Mao et al.. In
\medskip

\noindent$\bullet$ 2010, Babourova \& Frolov \cite{Babourova:2010wt}
came up with an alternative gravitational theory in which they claim
that orbital angular momentum can be a source of torsion. However,
their construct is inconsistent since already their Lagrangian
\cite{Babourova:2010wt}, Eq.\ (10), depends on the position ``vector''
and is as such no longer a covariant quantity. Still, in\medskip

\noindent$\bullet$ 2011, March et al.\ \cite{March:2011} reiterate Mao
et al., but take instead Mercury and Moon data and, later, data of the
Lageos satellite. Again they ``...make use of the autoparallel
trajectories, which in general differ {}from geodesics when torsion is
present.'' We saw already in the context of the Mao et al.\ discussion
that this leads to nowhere.

\section{Poincar\'e gauge theory, its general structure, quadratic
  gauge Lagrangians}
 
Let us now come back to the PG, which we only sketched in Sec.\
2. Consider the minimally coupled total Lagrangian $L$ in
(\ref{Lagr}). Since we want to leave the gravitational Lagrangian
$V=V(g_{\a\b},\vt^\a,T^\a,R^{\a\b})$ open for the time being, we
define the translational and Lorentz field excitations (or field
momenta),
\begin{eqnarray}\label{excit}
  {H_{\alpha}} = -\frac{{\partial}V}{{\partial}
    {T^{\alpha}}}\, , \quad {H_{{\alpha}{\beta}}} =
  -\frac{{\partial}V}{{\partial} {R^{{\alpha}{\beta}}}}\,.
\end{eqnarray}
As soon as we specify the explicit form of $V$, we can compute the H's
simply by partial differentiation.

The field equations of the PG read
\begin{eqnarray}\label{first}
  DH_{\alpha}- t_{\alpha}\hspace{6pt} & =& {\mathfrak{T}_{\alpha}} 
  \quad\hspace{7pt} (\text{{\sl First} grav.\ FEQ, }
  {\delta}/{\delta}{\vartheta}^{\alpha})\,,\\
  \label{second} D{H_{{\alpha}{\beta}}} - s_{{\alpha}{\beta}} &=&
  \mathfrak{S}_{{\alpha}{\beta}} \quad
  (\text{{\sl Second} grav.\ FEQ, }
  {\delta}/{\delta}{\Gamma}^{{\alpha}{\beta}})\,,\\
  {\delta L_{\text{mat}}}/{\delta\Psi} &=&
  0\qquad\hspace{1pt}\; (\text{{\sl Matter} FEQ, } {\delta/\delta\Psi})\,;
  \label{matter}
\end{eqnarray}
for their derivation see \cite{Reader}, Chapter 5, and the references
given there. The sources on the right-hand-side of {\sl First} and
{\sl Second} are the canonical energy-momentum
${\mathfrak{T}_{\alpha}}$ and spin $ \mathfrak{S}_{{\alpha}{\beta}}$
of {matter} defined in Eq.\ (\ref{sources}). The energy-momentum
and spin of the {\it gauge} fields are, respectively,
\begin{eqnarray}
  t_{\alpha}& := &e_{\alpha}
  \rfloor V + (e_{\alpha}\rfloor{T^{\beta}})\wedge {H_{\beta}}
  + (e_{\alpha}\rfloor
  {R^{{\beta}{\gamma}}})\wedge{H_{{\beta}{\gamma}}}\,, \\
  s_{\alpha}{}_{\beta}&:=& -
  \vartheta_{[\alpha}\wedge H_{\beta]}\,.
\end{eqnarray}
As we shall see later in detail, we will find the Einstein sector of
the PG if $H_\a=0$ and thus {\sl First} degenerates to the innocently
looking $-t_\a=\mathfrak{T}_\a$. 

Like in electrodynamics and in Yang-Mills theory, the gauge field
Lagrangian should be {\it algebraic} in the field strengths, here in
$T^\a$ and $R^{\a\b}$. Then we find second order partial differential
equations (PDEs) in the gauge variables $(\vt^\a,\Gamma^{\a\b})$.
Moreover, it should be {\it quadratic} in order to induce
quasi-linearity of the PDEs and thus wave type equations for both
gravitational field equations.

Symbolically, such a quadratic gauge Lagrangian, with the conventional
``weak'' gravitational constant $\kappa$, the ``strong'' gravitational
constant $\varrho$, and the cosmological constant $\Lambda_0$, reads
\begin{equation}\label{quadratic}
  V_{\text{qPG}}\sim  \frac 1\kappa  \left(R+X+\Lambda_0+\{T\}^{\!2}\,\right)
  +\frac{1}{\varrho}\{R \}^{\!2}\,.
\end{equation}
Here $R$ denotes the curvature scalar and $X\sim
\epsilon^{ijkl}R_{[ijkl]}$ the curvature pseudoscalar, $\{T\}^{\!2}$
symbolizes the sum of four torsion square pieces and $\{R \}^{\!2}$
the sum of eight curvature square pieces. The exact formula, which we
do not need here, can be found in Baekler et al.\ \cite{Baekler:2011jt}
or in \cite{Reader}, Eq.\ (5.13).

\section{Poincar\'e gauge theory and its viable subsets}

The quadratic gauge Lagrangian (\ref{quadratic}), taken together with
the field equations (\ref{first}), (\ref{second}), and (\ref{matter}),
represent the output of gauging the Poincar\'e group according to the
conventional gauge principles. This we consider to be the class of
standard torsion theories. To select a definite gauge Lagrangian
$V_{\text{qPG}}$ will be a task for the future, for details see
\cite{Reader}, Part B.

The simplest Lagrangian is the curvature scalar of the RC-spacetime
yielding the Einstein-Cartan theory of gravity (EC), see Figure 2:
\begin{eqnarray}\label{ECLagrangian}
  V_{\rm EC}&\sim&\nonumber \frac{1}{\kappa}\,R \sim\frac{1}{\kappa}
  \,^\star(\vt^\a\wedge\vt^\b)\wedge R_{\a\b}(\Gamma)\\
  &\sim& \frac{1}{\kappa}\,
  e_i{}^\a e_j{}^\b\,R^{ij}{}_{\alpha\beta}(\Gamma)\qquad (^\star\, =
  \text{Hodge dual})\,.
\end{eqnarray}

\begin{center}
\includegraphics[height=6cm]{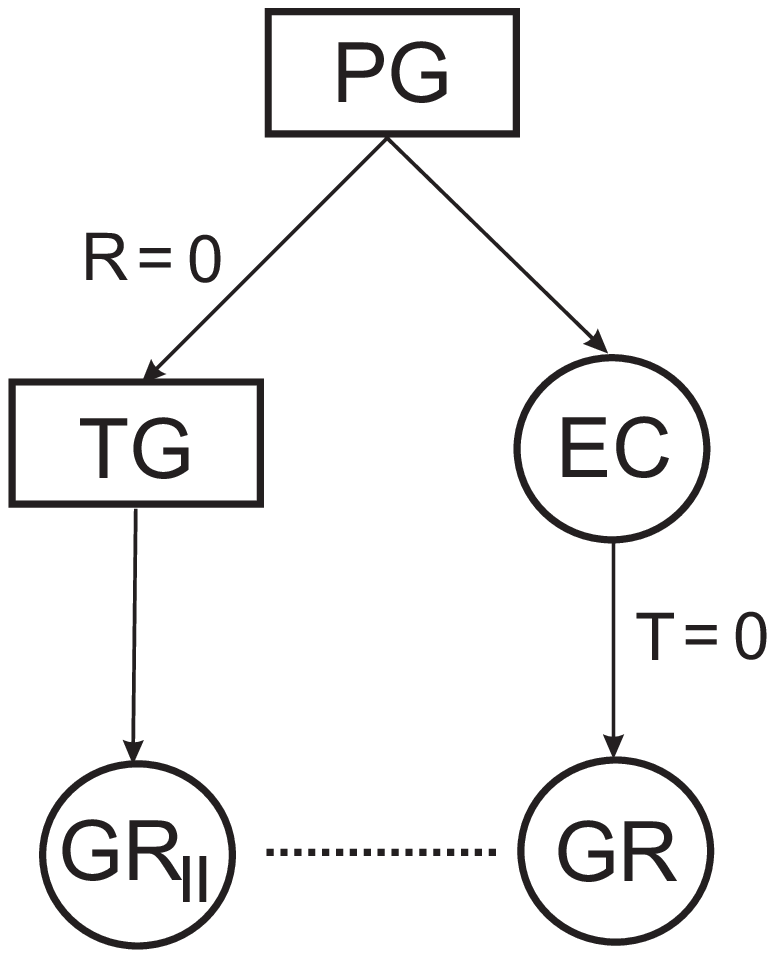}
\end{center}

\noindent{\it Fig,2.} Classification of Poincar\'e gauge theories of
gravity ({Blagojevi\'c, Hehl, Obukhov, see the frontispiece of
  \cite{Reader}): {\bf PG} = Poincar\'e gauge theory (of gravity),
{\bf EC} = Einstein-Cartan(-Sciama-Kibble) theory (of gravity), {\bf
  GR} = general relativity (Einstein's theory of gravity), {\bf TG} =
translation gauge theory (of gravity) aka teleparallel theory (of
gravity), {\bf GR}$_{||}$ = a specific TG known as teleparallel
equivalent of GR (spoken ``GR teleparallel''). The symbols in the
figure have the following meaning: rectangle $\Box$ $\rightarrow$
class of theories; circle ${\bigcirc}$ $\rightarrow$ definite viable
theories.

\bigskip

The EC can be put in such a form that it is represented by GR plus a
weak gravitational spin-spin contact interaction that only leads to
deviations {}from GR at extremely high matter densities. The critical
density at which GR breaks down is 
\begin{eqnarray}\label{critical}
 \rho_{\text{crit}}\sim 
 { m}/{\left(\lambda_{\text{Compton}}\ell_{\text{Planck}}^2
  \right)}\,.
\end{eqnarray} 
For a nucleon this is more than $10^{52}$ g/cm$^3$ or $10^{24}$ K,
with a critical length of $\ell_{\text{crit}}\sim 10^{-26}\,$cm. Thus,
EC is a viable gravitational theory. In cosmological applications it
could be of relevance. If in EC the material spin $\mathfrak{S}$ and
thus the torsion vanish, we recover GR, see Figure 2. Consequently, PG
contains the viable theories EC and GR and is as such a non-empty
framework with a reasonable GR limit.

Perhaps surprisingly, PG has a further subclass of a viable
gravitational theory. If we choose in the Lagrangian (\ref{quadratic})
the option with $\sim \frac {1}{\kappa}\,\{T\}^{\!2}$ and require the
vanishing of the curvature, that is, we imbed this Lagrangian in a
Weitzenb\"ock spacetime, then we have the Lagrangian of a translational
gauge theory of gravity (TG), namely
\begin{eqnarray}\label{translation}
 V_{\text{TG}}\sim  \frac {1}{\kappa}\,\{T\}^{\!2}      +
  R_{\alpha}{}^{\beta}\wedge \lambda^{\alpha}{}_{\beta}\,;
\end{eqnarray} 
here $\lambda^\a{}_\b$ is a Lagrange multiplier, see \cite{Reader},
Chapter 6, for details and literature.

Now we require additionally {\it local} Lorentz invariance and can single
out a definite version of $\{T\}^{\!2}$, namely
\begin{eqnarray} 
  V_{\text{GR$_{||}$}}&=&-\frac{1}{ 2\kappa}
  \,T^{\alpha}\!\wedge{}^{\star}\!\Bigl({ -}\,\underbrace{^{(1)}
    T_{\alpha}}_{\text{tensor}} + {
    2}\,\underbrace{^{(2)}T_{\alpha}}_{\text{vector}} + \frac 12\!
  \underbrace{^{(3)}T_{\alpha}}_{\text{axial
      vec.}}\Bigr)\nonumber\\
&& \hspace{2pt} +\,
  R_{\alpha}{}^{\beta}\wedge \lambda^{\alpha}{}_{\beta} \,,
\end{eqnarray}
the teleparallel equivalent of GR.  For scalar and for Maxwell matter,
that is, for $\mathfrak{T}_{ij}= \mathfrak{t}_{ij}$, it can be shown
that GR$_{||}$ and GR are equivalent, see \cite{Reader}, Chapter 6,
and as well the recent reviews of Aldrovandi \& Pereira
\cite{Aldrovandi:2013} and Maluf \cite{Maluf:2013gaa}. Hence we found
another viable version of PG, demonstrating the power of this
framework.

\section{The universally valid energy-momentum law in Poincar\'e gauge
  theory}\label{universal}
We generalize the corresponding law of GR, $D\mathfrak{t}_\a=0$,
together with the symmetry condition
$\vt_{[\a}\wedge\mathfrak{t}_{\b]}=0$, to the energy-momentum law and
angular momentum law of PG:
\begin{eqnarray}\label{energy1}
  D\mathfrak{T}_\a &\cong& (e_\a \rfloor T^\b)\wedge 
  \mathfrak{T}_\b+ (e_\a \rfloor R^{\b\g})\wedge
  \mathfrak{S}_{\b\g}\,,\\ \label{angular1}
  D\mathfrak{S}_{\a\b}&\cong&-\,\vt_{[\a}\wedge\mathfrak{T}_{\b]}\,.
\end{eqnarray}
These are the laws in exterior form calculus, for the tensor analysis
version, see (\ref{energy}) and (\ref{angular}). Incidentally, Cartan
assumed ad hoc that the right-hand-side of (\ref{energy1}) has to
vanish, similar as in GR's law $D\mathfrak{t}_\a=0$. However, if one
does the Noether ``algebra'' correctly, one finds (\ref{energy1}),
indeed; see, for instance, \cite{PRs}, Eq.\ (5.2.10) for $Q_{\a\b}=0$,
or \cite{Obukhov:2006gea}, Eq.\ (4.11). Interestingly enough, in 3
dimensions, Cartan's assumption turns out to be correct.

In order to isolate the torsion-dependent terms, we decompose the
connection $\Gamma^{\a\b}=\widetilde{\Gamma}^{\a\b}-K^{\a\b}$ into its
Riemannian part $\widetilde{\Gamma}^{\a\b}$ plus torsion-dependent
pieces. The contortion $K^{\a\b}$ is defined in (\ref{RCconn});
furthermore, we have $T^\a=K^\a{}_\b\wedge\vt^\b$. The split
connection will be substituted into the $D$ and $R^{\b\g}$ of
(\ref{energy1}). We find (${\cal L}=$ Lie derivative),
\begin{eqnarray}\label{energy2}\nonumber
&&  \widetilde{D}\left[ \mathfrak{T}_\a-{
      \mathfrak{S}^{\b\g}}(e_\a\rfloor K_{\b\g}) - {
      \mathfrak{S}^{\b\g}}\wedge ({\cal L}_{e_\a} K_{\b\g})\right]\\
&& \qquad\qquad \cong { \mathfrak{S}^{\b\g}}\wedge (e_\a\rfloor\widetilde{R}_{\b\g})\,.
\end{eqnarray}
This is a universally valid law for all PGs, independent of the
gravitational Lagrangian $V$. It shows conclusively that spin alone
does couple to contortion and hence to torsion. Note that its
right-hand-side depends only on the Riemannian piece, in contrast to
Eq.\ (\ref{energy}), which depends on the complete RC-curvature.

For vanishing torsion, we have $\widetilde{D}\mathfrak{T}_\a= {
  \mathfrak{S}^{\b\g}}\wedge (e_\a\rfloor \widetilde{R}_{\b\g})$, that
is, a momentum law that exhibits a spin-curvature force density on its
right-hand-side. This is the same structure as the
Mathisson-Papapetrou force in GR, with the difference that here the
force density does {\it not} contain integrated moments. Then,
$\widetilde{D}\mathfrak{t}_\a=0,$ with
$\mathfrak{t}_\a=\mathfrak{T}_\a -\widetilde{D}\mu_\a$ and $\mu_\a=-2
e_\b\rfloor \mathfrak{S}_\a{}^\b+\frac 12\vt_\a\wedge\left(e_\b\rfloor
  e_\g\rfloor\mathfrak{S}^{\b\g}\right)$.\medskip

In PG, integrate (\ref{energy2}) over a drop of a spin fluid. Then we
find the equation of motion of the momentum of the spin drop. It will
certainly not be an autoparallel curve, see \cite{DirkYuri:2008}.

\section{On nonminimal coupling}\label{nonminimal}

In this Letter we have confined our discussion to matter interacting
minimally with the gravitational field. However, currently
considerable attention is drawn to models with nonminimal
gravitational coupling see \cite{Bertolami,Nojiri,Puetzfeld}, for
example. In particular, in some theories such a generalized
interaction arises when the gravitational coupling constant is
replaced by a {\it coupling function} $F$ that depends on the
gravitational field strength (curvature and/or torsion).  The
corresponding modified Lagrangian reads $F\,L_{\rm mat}$. One can
verify that the metrical energy-momentum tensor
$\sqrt{-g}\,\mathfrak{t}_{ij} = 2\delta (\sqrt{-g}L_{\rm mat})/\delta
g^{ij}$ even for the spinless matter is not covariantly conserved
\cite{Koivisto,Bertolami}, but instead it satisfies the balance law
\begin{eqnarray}\label{conservation}
  \nabla^i \mathfrak{t}_{ij} = -\left( g_{ij} L_{\rm mat} 
    + \mathfrak{t}_{ij} \right) \frac{ \nabla^iF}{F}\,.
\end{eqnarray}  
As an immediate consequence, we find that the motion of a test particle 
or body is non-geodetic since an extra force is acting on it, which is 
proportional to the gradient of the coupling function $F$. 

It was demonstrated in \cite{Puetzfeld} that the modified conservation
law (\ref{conservation}) leads to non-geodetic motion, in which $F$
can depend arbitrarily on the components of the Riemann curvature
tensor (in practice, being a scalar, $F$ is any function of all
possible algebraic invariants built {}from the curvature).  Furthermore,
in \cite{ObukhovPuetzfeld:2013} it was shown that the remarkably
simple law (\ref{conservation}) is also true for the coupling function
$F$ that depends arbitrarily on the Riemann-Cartan curvature {\it and}
torsion tensors.

This implies that a nonminimal coupling of matter to gravity
represents actually a loophole that allows to detect the possible
non-Riemannian structure of spacetime by means of spinless matter. 
Such a loophole, however, is qualitatively different {}from the 
``Hypothesis T'' which we discuss in the next section. 

\section{Our answer to Mao et al.\ and March et al.}\label{answerMao}

In a discussion, Tegmark made an attempt to elucidate the Mao et al.\
philosophy by defining the {\it Hypothesis T\/} \cite{Tegmark}:
``{There's a consistent nonstandard gravity theory where torsion
  couples to macroscopic rotation}''.  Taking this hypothesis as their
starting point, Mao et al.\ believe that they could constrain the
possible torsion of spacetime in such a nonstandard gravity theory via
the experimental results of GPB.

If we take PG as the standard torsion theory, then rotating quartz
balls are blind to torsion, as we underlined again in
Sec.\ \ref{universal}. Even if we do not commit ourselves to a specific
form of the gravitational Lagrangian $V$, the energy-momentum
law commands that torsion only couples to spin, see equation
(\ref{energy2}) and the two terms that couple spin ${\cal S}$
directly to the contortion $K$.

Moreover, as Flanagan \& Rosenthal \cite{Flanagan:2007dc} and two of
us \cite{DirkYuri:2008} have shown, the teleparallel theory used by
Mao et al.\ also cannot provide a framework for measuring
torsion. Hence Hypothesis T is empty so far. Furthermore, Mao et al.\
postulated the autoparallel as an equation of motion, which is
incorrect in a general-relativistic set-up anyway, quite independent
of the field equations.
 
It is a general rule in physics that one can only measure a certain
quantity provided one has a consistent theory about this quantity in
the first place. If one wants to measure, say, an acceleration of a
particle, one has first to define an acceleration via $\vec{a}=d^2
\vec{x}/dt^2$. After these considerations, one can measure
$\vec{a}$. {\it A sensible interpretation of experiments in physics
  usually requires preceding theoretical groundwork.} By the same
token, we have first to develop a consistent theory of the torsion
before we are able to measure it.

But there is even a more direct argument that is lethal to Hypothesis
T: In the whole of the Mao et al.\ paper, the authors only speak about
the {\it vacuum} field equations. They never address the question of
how a field equation could look like where ``macroscopic rotation'',
that is, macroscopic orbital angular momentum features as a source of
a gravitational field equation. Mao et al.\ claim that in other papers
there is an ``assumption that orbital angular momentum cannot be the
source of torsion.'' In fact, orbital angular momentum as a tensor is
only known for {\it extended} structures, never as a density existing
at one point. It is a quantity alien to local field theory. The
torsion tensor, however, is a local point-dependent
object. Accordingly, in a {\it local} field equation, these two
quantities cannot be related to each other. That is, orbital angular
momentum cannot be the source of torsion on account of the different
nature of those two objects.

In special relativity in {\it Cartesian} coordinates, the orbital
angular momentum flux density is $x_{[i}{\cal  T}_{j]}{}^k$,
with the position vector ${x_i}$ and the energy-momentum tensor ${\cal 
  T}_{i}{}^j$. The divergence of the orbital angular momentum flux
density reads
\begin{eqnarray}\partial_k(x_{[i}{\cal  T}_{j]}{}^k )=
  {\cal  T}_{[ji]}+x_{[i|}\partial_{k}{\cal  T}_{|j]}{}^k\,.
\end{eqnarray} 
If the action of a physical system without (intrinsic) spin angular
momentum is invariant under spacetime translations and (Lorentz)
rotations, the energy-momentum is conserved, $\partial_k{\cal
  T}_{i}{}^{\,k}=0$, and we find the angular momentum conservation law
\begin{eqnarray}
\partial_k(x_{[i}{\cal  T}_{j]}{}^k ) = {\cal  T}_{[ji]} = 0\,.
\end{eqnarray}

The orbital angular momentum $x_{[i}{\cal T}_{j]}{}^k$ is {\it not} a
tensor in curvilinear coordinates. In the words of Truesdell
\cite{Truesdell:1964}, it is a quantity that is ``{\it not}
indifferent'' to coordinate transformations, whereas legitimate
field-theoretical quantities should be indifferent.  Accordingly, this
quantity does not exist as a {\it local} quantity in a curved (in
particular, in Riemann-Cartan) spacetime. However, the conservation
law of the angular momentum, namely $ {\cal T}_{[ji]} = 0$, can be
generalized to Riemann-Cartan spacetimes, provided the momentum law is
fulfilled. When, in addition, matter possesses the {\it intrinsic} spin
angular momentum ${\cal S}_{ij}{}^k$, it contributes to the balance
law the divergence $D_k{\cal S}_{ij}{}^k$ thus providing the total
angular momentum law $D_k{\cal S}_{ij}{}^k - {\cal T}_{[ij]}=0$. The
latter is meaningful even if the orbital angular momentum does not
exist in curved space: In contrast to its orbital partner, spin
angular momentum ${\cal S}_{ij}{}^{k}$ is a well-defined ``indifferent''
tensor in a Riemann-Cartan spacetime and can act as a source.

Hypothesis T is untenable, since it links the field theoretical notion
``torsion'' with the orbital angular momentum of an extended
structure; this net orbital angular momentum cannot be represented as
an integral over a local orbital angular momentum {\it density}, since
such a density does not exist. We thus conclude that Hypothesis T is
empty.

No counterexample is known to our result. Our conclusion can be found
in the last phrase of our Abstract.

Eventually, we have also an optimistic message: Already in 1983, Ni
\cite{Wei-Tou:1983} suggested to build gyroscopes with spin-polarized
balls as active elements consisting of solid helium-three ($^3$He) and
to put them into orbit around the Earth. Ni has also used for experiments
in the gravitational field dysprosium-iron compounds Dy$_6$Fe$_{23}$,
see \cite{Wei-Tou:2010}, with a relatively high net spin of about
$0.4$ electron spins per atom, but with no disturbing magnetic
moment. With such tools one could hope to find torsion, if it exists
in nature.

\medskip

\noindent{\bf Acknowledgments}  

One of us (F.W.H.) is grateful to Max Tegmark and also to Yi Mao for
an extended and lively email discussion; even though we could not
agree on a joint position, this discussion was very helpful. Likewise
we would like to thank Giovanni Bellettini and Riccardo March for
several discussions. We thank Milutin Blagojevi\'c for the drawing of
Figure 2. Moreover, we thank Wei-Tou Ni, Neil Russell, Helmut Rumpf,
and Francis Everitt for useful remarks. F.W.H.\ was partially
supported by the German--Israeli Foundation for Scientific Research
and Development (GIF), Research Grant No.\ 1078--107.14/2009. D.P.\
was supported by the Deutsche Forschungsgemeinschaft (DFG) through the
grant LA-905/8-1.

\begin{footnotesize}
 
\end{footnotesize}

\end{document}